\documentclass[aps,twocolumn,pra,preprintnumbers,showpacs,tightenlines]{revtex4}
\usepackage{amsmath,graphicx,bbm,mathrsfs,amssymb,pst-all,bm,color}

\input{tcilatex}
\begin{document}

\title{Robust and scalable optical one-way quantum computation}
\author{Hefeng Wang, Chui-Ping Yang, and Franco Nori}
\affiliation{Advanced Science Institute, RIKEN, Wako-shi, Saitama,
351-0198, Japan and Physics Department, The University of
Michigan, Ann Arbor, Michigan 48109-1040, USA}

\begin{abstract}
We propose an efficient approach for deterministically generating scalable
cluster states with photons. This approach involves unitary transformations
performed on atoms coupled to optical cavities. Its operation cost scales
linearly with the number of qubits in the cluster state, and photon qubits
are encoded such that single-qubit operations can be easily implemented by
using linear optics. Robust optical one-way quantum computation can be
performed since cluster states can be stored in atoms and then transferred
to photons that can be easily operated and measured. Therefore, this
proposal could help in performing robust large-scale optical one-way quantum
computation.

\noindent
\end{abstract}

\pacs{42.50.Ex, 03.67.Lx}
\maketitle

\section{Introduction}

\label{intro}

In principle, linear optical elements~(e.g., beam splitters, phase
shifters, etc.), combined with single-photon sources and detectors
can be used for efficient quantum information
processing~\cite{klm}. Experimental progress in optical systems
has demonstrated control of photonic qubits, quantum gates, and
small quantum algorithms~(e.g.,\cite{kok, obr, wal, pre}). Optical
quantum computation~(QC) has been suggested~\cite{nie, brow} using
cluster states~\cite{br1, rau1, br2, tan1, you1, wxb, tan2}.
One-way optical QC
using a four-photon cluster state has been demonstrated experimentally~\cite%
{wal, pre, tame}. In spite of this progress, scalable optical one-way QC
still remains elusive because of the difficulty in generating cluster states
with a large number of qubits.

Photon cluster states are excellent candidates for one-way QC because of the
fast and easy implementation of single-qubit operations on photons, and also
because photonic qubits have negligible decoherence. However, it is
difficult to generate cluster states with photons because of the absence of
significant interactions between photons. In general, there are two types of
method for generating cluster states with photons: ($1$)~by introducing an
effective interaction between photons through measurements~\cite{nie, brow};
($2$)~by using a nonlinear optical process, parametric down-conversion~\cite%
{pre, wal, val, park, chen}. Generation of cluster states through
measurements~\cite{klm} is probabilistic. Because of its intrinsically
probabilistic nature, the method based on down-conversion is exponentially
inefficient for generating large cluster states~\cite{wal}, therefore it can
only prepare cluster states of a few qubits. In short, both of these methods
are not efficient in generating cluster states with a large number of photon
qubits.

In this paper, we propose an approach that is different from the previously
proposed methods for generating cluster states with photons. This method is
deterministic and efficient in generating scalable photon cluster states.
The cost of the approach scales linearly with the number of qubits in a
cluster state. The standard encoding of photon qubits allows
easy-to-implement single-qubit operations using passive linear optics.

Our approach for generating cluster states with photons is as follows:
first, generate a cluster state in atoms trapped in the periodic potential
of an optical lattice, then transfer the cluster state from the atomic
system to photons through the coupling between the atoms and optical
cavities. Note that atomic cluster states can be easily obtained in
experiments~\cite{man, blo} on optical lattices through next-neighbor
interactions, it can be achieved in a \textit{single} operational step and
is \textit{independent} of the size of the systems~\cite{man}.

For robust and scalable one-way QC, large cluster states need to be
generated and stored for performing single-qubit operations and readout.
Cluster states can be easily \textit{generated and stored with atoms}, but
it is difficult to perform measurements on atoms. In contrast, it is easy to
perform \textit{measurements} on photons, but it is difficult to store
quantum states using photons. Thus, this hybrid proposal uses the best from
atomic and photonic qubits, to provide robust one-way QC. Namely, to
generate and store cluster states in an atomic system, then transfer to
photons the states that are subjected to measurements, and then perform
single-qubit operations and measurements on photonic qubits.

In this work we focus on transferring to a photonic system a cluster state
originally generated in an atomic system. As discussed above, this is a
crucial step in our proposal. The method presented below employs five-level
atoms coupled to optical cavities. This method has the following advantages:
$(1)$~Neither the cavity-mode frequencies nor the atomic level spacings need
to be adjusted during the operation process; $(2)$~No measurement is
required. Our approach for generating photon cluster states is based on
unitary transformations, i.e., a deterministic method; $(3)$~There is no
time limitation for moving atoms in or out of the cavities, therefore it
should be relatively easy to manipulate the system in experiments; $(4)$~We
choose the traditional encoding of a photonic qubit in each cavity, by using
a photon in either a left-circularly polarized mode or a right-circularly
polarized mode of the cavity. With this encoding, single-qubit operations
are easy to implement by using polarization rotators~\cite{klm}. Even though
we consider here natural atoms, in the future, this proposal could be
extended to artificial atoms~\cite{you, nori, bul, bul1}.

The structure of this work is as follows: In Secs.~\ref{trans1} and \ref%
{trans2}, we explain how to generate cluster states on photonic systems
using five-, four- and three-level atoms interacting with cavities,
respectively. We close with a conclusion.

\section{Transferring a cluster state from atoms in an optical lattice to
polarized photons: Five-level atomic system}

\label{trans1}

Consider a system composed of $n$ atoms and $n$ cavities. Each atom has a
five-level structure as depicted in Fig.~$1$, and is placed in a two-mode
cavity. The two modes in each cavity are left-circularly polarized~($\sigma
_{L}^{-}$) and right-circularly polarized~($\sigma _{R}^{+}$), respectively.
For each atom, the two lowest energy levels $|g\rangle $ and $|g^{\prime
}\rangle $ represent the two logic states of a qubit; while for each cavity,
the two logic states of a qubit are represented by the occupation of a
photon~(sub-index \textit{p}) in the left- and right-circularly polarized
modes of a cavity as:
\begin{equation}
|0\rangle _{p}=|0\rangle _{L}|1\rangle _{R},\text{ }|1\rangle _{p}=|1\rangle
_{L}|0\rangle _{R},
\end{equation}%
where $|k\rangle _{L}|m\rangle _{R}$ represents the state of the cavity with
$k$ or $m$ photons in the left- or right-circularly polarized modes.

Let us assume that an $n$-qubit cluster state $|\Psi _{C}\rangle $ was
prepared in the $n$-atom system, and each two-mode cavity is in the vacuum
state $|\emptyset \rangle _{p}=|0\rangle _{L}|0\rangle _{R}$. Thus, the
initial state of the whole system is $|\Psi _{C}\rangle _{a}\otimes
|\emptyset \rangle _{p}^{\otimes n}$, where the subscripts $a$ and $p$
represent the atomic and the photonic systems, respectively. The task is to
perform the state transfer
\begin{equation}
|\Psi _{C}\rangle _{a}|\emptyset \rangle _{p}^{\otimes n}\text{ \ }%
\longrightarrow \text{ \ }|g\rangle _{a}^{\otimes n}|\Psi _{C}\rangle _{p},
\end{equation}%
i.e., transfer the cluster state from the atomic system to the photons
inside the optical cavities, then photons leaking out of the cavities would
be in the same cluster state as originally prepared in the atomic system.

Alternatively, we can also transfer an atomic single-qubit state of the
cluster state to a photonic qubit~(see discussion section below). This state
transfer can be achieved through the following transformation~(performed on
each two-mode cavity with an atom inside)
\begin{eqnarray}
|g\rangle |0\rangle _{L}|0\rangle _{R} &\longrightarrow &|g\rangle |0\rangle
_{L}|1\rangle _{R},  \notag \\
|g^{\prime }\rangle |0\rangle _{L}|0\rangle _{R} &\longrightarrow &|g\rangle
|1\rangle _{L}|0\rangle _{R}.
\end{eqnarray}%
Below, we will show the procedure to implement this transformation.
\begin{figure}[tbp]
\includegraphics[width=7.5cm, height=4cm]{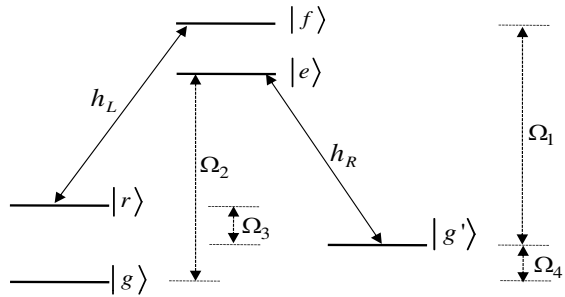}
\caption{Energy diagram of an atom with five levels. The transitions $%
|g^{\prime }\rangle \leftrightarrow |e\rangle $ and $|r\rangle
\leftrightarrow |f\rangle $ are resonantly coupled to the left~($\protect%
\sigma _{L}^{-}$) and right~($\protect\sigma _{R}^{+}$) circularly polarized
mode of the cavity with coupling strengths $h_{L}$ and $h_{R}$. The
transitions $|g^{\prime }\rangle \leftrightarrow |f\rangle ,|g\rangle
\leftrightarrow |e\rangle ,|g^{\prime }\rangle \leftrightarrow |r\rangle $,
and $|g\rangle \leftrightarrow |g^{\prime }\rangle $ are driven by laser
fields with Rabi frequencies $\Omega _{1},\Omega _{2},\Omega _{3}$, and $%
\Omega _{4}$, respectively.}
\end{figure}

\subsection{Transformation of a state from atoms to polarized photons}

To achieve this transformation, two interactions between the atom with the
two cavity modes are needed. One is the resonant interaction between the $%
|r\rangle \leftrightarrow |f\rangle $ transition and the $\sigma _{L}^{-}$
mode. The other one is the resonant interaction between the $|g^{\prime
}\rangle \leftrightarrow |e\rangle $ transition and the $\sigma _{R}^{+}$
mode. In the interaction picture, the Hamiltonians for these are
\begin{equation}
H_{L}=h_{L}\left( a_{L}|f\rangle \langle r|+\text{H.c.}\right) ,
\end{equation}%
and
\begin{equation}
H_{R}=h_{R}\left( a_{R}|e\rangle \langle g^{\prime }|+\text{H.c.}\right) ,
\end{equation}%
\noindent where $h_{L}$~($h_{R}$) is the coupling strength of the atom with
the $\sigma _{L}^{-}$~($\sigma _{R}^{+}$) mode of the cavity; $a_{L}$~($%
a_{R} $) is the annihilation operator of the $\sigma _{L}^{-}$~($\sigma
_{R}^{+}$) mode of the cavity.

The transformation in Eq.~($3$) can be achieved in four steps as follows:

\textit{Step~(i):} Apply a pulse to the atom for a time interval $\tau _{1}$%
, which is resonant with the $|g^{\prime }\rangle \leftrightarrow |f\rangle $
transition; then wait for a time interval $\tau _{L}$. We denote $\Omega
_{1} $ the Rabi frequency of the pulse and $\phi _{1}$ the phase of the
pulse. The time evolution for this step is as follows:

\textit{(ia)} During the time interval $\tau _{1}$, the pulse applied to the
atom leads to the transformation
\begin{eqnarray}
|g^{\prime }\rangle |0\rangle _{L}|0\rangle _{R} &\rightarrow &\left[ \cos (%
\frac{\Omega _{1}}{2}\tau _{1})|g^{\prime }\rangle -ie^{-i\phi _{1}}\sin (%
\frac{\Omega _{1}}{2}\tau _{1})|f\rangle \right]  \notag \\
&&\otimes |0\rangle _{L}|0\rangle _{R}.
\end{eqnarray}%
For $\frac{\Omega _{1}}{2}\tau _{1}=\pi /2$, $\phi _{1}=-\pi /2$, the state $%
|g^{\prime }\rangle |0\rangle _{L}|0\rangle _{R}$\ is transformed into $%
|f\rangle |0\rangle _{L}|0\rangle _{R}$.

\textit{(ib)} During the waiting time $\tau _{L}$, the $|r\rangle
\leftrightarrow |f\rangle $ transition of the atom is resonant with the $%
\sigma _{L}^{-}$\ mode of the cavity. The Hamiltonian describing this
process is $H_{L}$ in Eq.~($4$). The operator $U_{L}=\exp \left( -iH_{L}\tau
_{L}\right) $ performs the transformation
\begin{eqnarray}
|f\rangle |0\rangle _{L}|0\rangle _{R} &\rightarrow &\cos (h_{L}\tau
_{L})|f\rangle |0\rangle _{L}|0\rangle _{R}  \notag \\
&&-i\sin (h_{L}\tau _{L})|r\rangle |1\rangle _{L}|0\rangle _{R}.
\end{eqnarray}%
For $h_{L}\tau _{L}=\pi /2$, the state $|f\rangle |0\rangle _{L}|0\rangle
_{R}$ evolves to $-i|r\rangle |1\rangle _{L}|0\rangle _{R}$.

\textit{Step~(ii):} Apply a pulse to the atom for a duration $\tau _{2}$.
The pulse applied to the atom is resonant with the $|g^{\prime }\rangle
\leftrightarrow |r\rangle $ transition. This process leads to the
transformation%
\begin{eqnarray}
-i|r\rangle |1\rangle _{L}|0\rangle _{R} &\rightarrow &-i\left[ \cos (\frac{%
\Omega _{2}}{2}\tau _{2})|r\rangle -ie^{i\phi _{2}}\sin (\frac{\Omega _{2}}{2%
}\tau _{2})|g^{\prime }\rangle \right]  \notag \\
&&\otimes |1\rangle _{L}|0\rangle _{R}.
\end{eqnarray}%
With $\frac{\Omega _{2}}{2}\tau _{2}=\pi /2$ and $\phi _{2}=\pi $, the state
$-i|r\rangle |1\rangle _{L}|0\rangle _{R}$ is transformed to $|g^{\prime
}\rangle |1\rangle _{L}|0\rangle _{R}$.

\textit{Step~(iii):} Apply a pulse to the atom for a time interval $\tau
_{3} $, which is resonant with the $|g\rangle \leftrightarrow |e\rangle $
transition; then wait for a time interval $\tau _{R}$. The time evolution
for this step is as follows:

\textit{(iiia)} During the time interval $\tau _{3}$, the pulse applied to
the atom leads to the transformation%
\begin{eqnarray}
|g\rangle |0\rangle _{L}|0\rangle _{R} &\rightarrow &\left[ \cos (\frac{%
\Omega _{3}}{2}\tau _{3})|g\rangle -ie^{-i\phi _{3}}\sin (\frac{\Omega _{3}}{%
2}\tau _{3})|e\rangle \right]  \notag \\
&&\otimes |0\rangle _{L}|0\rangle _{R}.
\end{eqnarray}%
For $\frac{\Omega _{3}}{2}\tau _{3}=\pi /2$ and $\phi _{3}=\pi $, the state $%
|g\rangle |0\rangle _{L}|0\rangle _{R}$\ becomes $i|e\rangle |0\rangle
_{L}|0\rangle _{R}$.

\textit{(iiib)} During the waiting time $\tau _{R}$, the $|g^{\prime
}\rangle \leftrightarrow |e\rangle $ transition of the atom is resonant with
the $\sigma _{R}^{+}$\ mode of the cavity. The Hamiltonian for this process
is $H_{R}$ in Eq.~($5$). Then $U_{R}=\exp \left( -iH_{R}\tau _{R}\right) $
transforms%
\begin{eqnarray}
i|e\rangle |0\rangle _{L}|0\rangle _{R} &\rightarrow &i\cos (h_{R}\tau
_{R})|e\rangle |0\rangle _{L}|0\rangle _{R}  \notag \\
&&+\sin (h_{R}\tau _{R})|g^{\prime }\rangle |0\rangle _{L}|1\rangle _{R}.
\end{eqnarray}%
With $h_{R}\tau _{R}=\pi /2$, the state $i|e\rangle |0\rangle _{L}|0\rangle
_{R}$ becomes $|g^{\prime }\rangle |0\rangle _{L}|1\rangle _{R}$.

\textit{Step~(iv):} Apply a pulse to the atom for a time interval $\tau _{4}$%
. The pulse is resonant with the $|g^{\prime }\rangle \leftrightarrow
|g\rangle $ transition. Thus we have the transformations%
\begin{eqnarray}
|g^{\prime }\rangle |1\rangle _{L}|0\rangle _{R} &\rightarrow &\left[ \cos (%
\frac{\Omega _{4}}{2}\tau _{4})|g^{\prime }\rangle -ie^{i\phi _{4}}\sin (%
\frac{\Omega 4}{2}\tau _{4})|g\rangle \right]  \notag \\
&&\otimes |1\rangle _{L}|0\rangle _{R}.
\end{eqnarray}%
and
\begin{eqnarray}
|g^{\prime }\rangle |0\rangle _{L}|1\rangle _{R} &\rightarrow &\left[ \cos (%
\frac{\Omega _{4}}{2}\tau _{4})|g^{\prime }\rangle -ie^{i\phi _{4}}\sin (%
\frac{\Omega 4}{2}\tau _{4})|g\rangle \right]  \notag \\
&&\otimes |0\rangle _{L}|1\rangle _{R}.
\end{eqnarray}%
With $\frac{\Omega _{4}}{2}\tau _{4}=\pi /2$ and $\phi _{4}=\pi /2$, the
state $|g^{\prime }\rangle |1\rangle _{L}|0\rangle _{R}$ is transformed to $%
|g\rangle |1\rangle _{L}|0\rangle _{R}$, and the state $|g^{\prime }\rangle
|0\rangle _{L}|1\rangle _{R}$ becomes $|g\rangle |0\rangle _{L}|1\rangle
_{R} $.

After this operation, the atom is decoupled from the cavity, and is in a
stable state. One can easily check that the transformation in Eq.~($3$) is
achieved in the four steps above.

\subsection{Fidelity of the transformation}

Let us now study the fidelity of the state transfer operation described
above. We assume that the pulses applied to the atoms can be controlled
within a very short time~(e.g., by increasing the pulse amplitude), such
that the dissipation of the system during the pulse is negligibly small. In
this case, the dissipation of the system would appear in the time evolution
operations in step~(\textit{ib}) and step~(\textit{iiib}). Before any photon
leaks out of each cavity, the Hamiltonians $H_{L}$ and $H_{R}$ become%
\begin{equation}
H_{L}^{\prime }=h_{L}\left( a_{L}|f\rangle \langle r|+\text{H.c.}\right)
-i\gamma _{f}|f\rangle \langle f|-i\kappa _{L}a_{L}^{\dagger }a_{L},
\end{equation}%
\begin{equation}
H_{R}^{\prime }=h_{R}\left( a_{R}|e\rangle \langle g^{\prime }|+\text{H.c.}%
\right) -i\gamma _{e}|e\rangle \langle e|-i\kappa _{R}a_{R}^{\dagger }a_{R},
\end{equation}%
where $\gamma _{f}$ ($\gamma _{e}$) is the decay rate of the atomic level $%
|f\rangle $\ ($|e\rangle $), and $\kappa _{L}$ ($\kappa _{R}$)\ is the decay
rate of the $\sigma _{L}^{-}$\ ($\sigma _{R}^{+}$) mode of the cavity. For
simplicity, we assume $h_{R}=h$, $h_{L}=s\cdot h$, where $s>1$, $\gamma
_{e}=\gamma _{f}=\gamma $, and $\kappa _{L}=\kappa _{R}=\kappa $.

We now numerically calculate the evolution of the system governed by the
Hamiltonians above. The quality of the state transfer process in Eq.~($3$)
can be described by the fidelity of the state transfer operation%
\begin{equation}
F=Tr\left[ \left( \rho _{p}^{\mathrm{id}}\right) ^{1/2}\rho _{p}\left( \rho
_{p}^{\mathrm{id}}\right) ^{1/2}\right] ^{1/2},
\end{equation}%
where $\rho _{p}$ represents the photon~($p$) temporal reduced density
matrix, after tracing over the states of the atom, and $\rho _{p}^{\mathrm{id%
}}$ represents the reduced density matrix in the ideal~($=$ id) case without
considering the dissipation of the system. From Eqs.~($13,14$), one can see
that the atomic spontaneous decay plays the same role as the cavity decay in
the dissipation process. They both have the same effect on the fidelity, and
thus the fidelity of the state transfer operation can be improved by
choosing atoms that have a long energy relaxation time or by improving the
coupling between the cavity and the atom. Assume $s=1.2$, then $%
h_{L}=1.2h_{R}=1.2h$, and the fidelity\ versus the spontaneous atomic decay
rate $\gamma /h$ for different cavity decay rates $\kappa $ is shown in Fig.~%
$2$. Cluster states with a high fidelity can be obtained efficiently using
existing multi-particle entanglement purification protocols~\cite{dur}.
\begin{figure}[tbp]
\includegraphics[width=7.5cm, height=5cm]{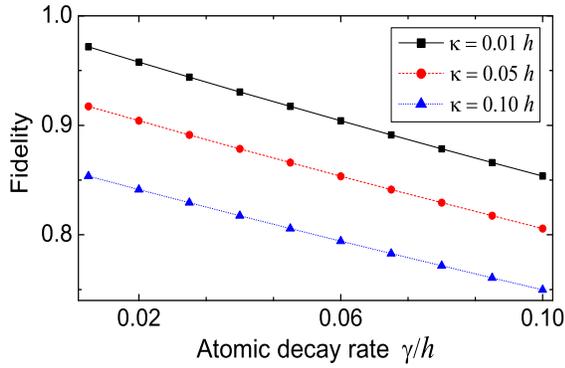}
\caption{(Color online) Fidelity~($F$) of the transformation in Eq.~($3$)
implemented on a five-level atom coupled with a two-mode cavity versus the
spontaneous atomic decay rate $\protect\gamma /h$. The solid, dashed and
dotted lines refer to the cavity decay rates $\protect\kappa %
=0.1h,0.05h,0.01h$, respectively.}
\end{figure}

\subsection{Linear scaling}

As shown above, four steps are needed to transfer the state from an atomic
qubit to a photonic qubit. Therefore an $n$-qubit photon cluster state can
be created in $4n$ steps. In other words, the operation cost scales \textit{%
linearly} with the number of qubits in the cluster state. Hence, this
approach could provide an efficient generation of scalable cluster states
with photons.

\subsection{Atomic candidate}

As a possible implementation, the $^{87}$Rb atom can be used as the
five-level atom. The atomic levels $|g\rangle $, $|g^{\prime }\rangle $ and $%
|r\rangle $ are $|F=1,m=-1\rangle $, $|F=1,m=0\rangle $ and $|F=2,m=0\rangle
$ of $5^{2}$S$_{1/2}$, respectively; $|e\rangle $ and $|f\rangle $, are $%
|F=1,m=1\rangle $ and $|F=2,m=-1\rangle $ of $5^{2}$P$_{1/2}$, respectively.

To generate an $n$-qubit photon cluster state, we need to send $n$ atoms
that encode the cluster state into $n$ two-mode cavities. Atomic cluster
states can be easily created on optical lattices~\cite{man}, and it is also
possible to load atoms into cavities through transverse optical lattice
potentials~\cite{duan, bei, sau}. The process of transferring the atomic
cluster states to the photonic qubits must be completed in a very short
time. To do this, one can prepare an array of cavities and load the atoms
into the cavities through transverse optical lattice potentials. After a
certain time, the photons that leak out of the cavities are in the same
cluster state as the cluster state originally prepared in the atomic system.
This process is illustrated in Fig.~$3(a)$.
\begin{figure}[tbp]
\includegraphics[width=7.0cm, height=7.5cm]{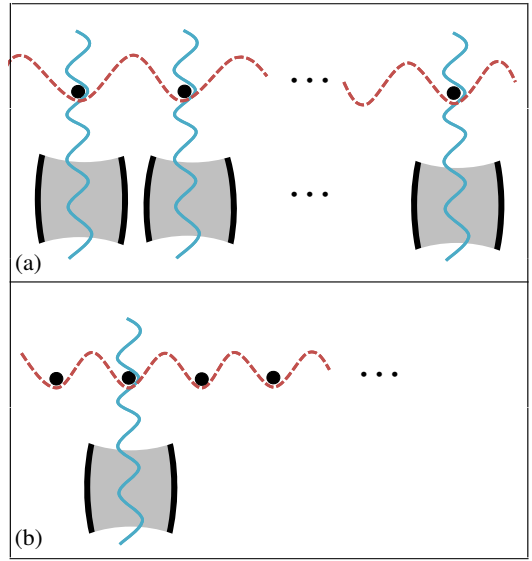}
\caption{(Color online) (a)~Schematic diagram of the setup for translating
atoms in optical lattices into an array of cavities through transverse
optical lattice potentials. (b)~Setup for translating an atom in an optical
lattice that is subjected to measurement into a cavity through a transverse
optical lattice potential. There is no order requirement for loading atoms
into the cavity using this setup.}
\end{figure}
\newline

\section{Transferring a cluster state from atoms in an optical lattice to
polarized photons: Four- and three-level atomic systems}

\label{trans2}

As shown above, five-level atoms can be employed in performing the
transformation in Eq.~($3$). We note that the transfer of cluster states
from atomic to photonic systems can also be realized by using four-level
atoms coupled to cavities. Suppose that $n$ two-mode cavities are initially
prepared in the state $\left( |0\rangle _{p}\right) ^{\otimes n}$. To
transfer the cluster state $|\Psi _{C}\rangle $ from an atomic system to
photons, one needs to perform the transformation%
\begin{equation}
|\Psi _{C}\rangle _{a}\left( |0\rangle _{p}\right) ^{\otimes n}\text{ \ }%
\longrightarrow \text{ \ }|g\rangle _{a}^{\otimes n}|\Psi _{C}\rangle _{p}.
\end{equation}%
This process can be done by applying $n$ swap gates between the two coupled
systems. Each swap gate acting on an atom in a two-mode cavity performs the
transformation $|g\rangle _{a}|0\rangle _{p}\rightarrow |g\rangle
_{a}|0\rangle _{p}$, $|g^{\prime }\rangle _{a}|0\rangle _{p}\rightarrow
|g\rangle _{a}|1\rangle _{p}$, i.e.,
\begin{eqnarray}
|g\rangle |0\rangle _{L}|1\rangle _{R} &\longrightarrow &|g\rangle |0\rangle
_{L}|1\rangle _{R},  \notag \\
|g^{\prime }\rangle |0\rangle _{L}|1\rangle _{R} &\longrightarrow &|g\rangle
|1\rangle _{L}|0\rangle _{R}.
\end{eqnarray}

In this case, the four energy levels of the atom are shown in Fig.~$4$. The
swap operations described by Eq.~($17$) can be implemented in three steps:

\textit{Step~(i)}: Let the system evolve a time $\tau _{1}$ under $%
H_{1}=h_{R}\left( a_{R}|f\rangle \langle g^{\prime }|+\text{H.c.}\right) $.
The operator $U_{1}=\exp \left( -iH_{1}\tau _{1}\right) $ performs the
transformation
\begin{eqnarray}
|g^{\prime }\rangle |0\rangle _{L}|1\rangle _{R} &\rightarrow &\cos
(h_{R}\tau _{1})|g^{\prime }\rangle |0\rangle _{L}|1\rangle _{R}  \notag \\
&&-i\sin (h_{R}\tau _{1})|f\rangle |0\rangle _{L}|0\rangle _{R}.
\end{eqnarray}%
With $h_{R}\tau _{1}=\pi /2$, the state $|g^{\prime }\rangle |0\rangle
_{L}|1\rangle _{R}$ evolves to $-i|f\rangle |0\rangle _{L}|0\rangle _{R}$.

\textit{Step~(ii)}: Apply a pulse to the atom for a time interval $\tau _{2}$
resonant with $|f\rangle \leftrightarrow |r\rangle $ transition. This
process leads to the transformation
\begin{eqnarray}
-i|f\rangle |0\rangle _{L}|0\rangle _{R} &\rightarrow &-i\left[ \cos (\frac{%
\Omega _{1}}{2}\tau _{2})|f\rangle -ie^{i\phi _{1}}\sin (\frac{\Omega _{1}}{2%
}\tau _{2})|r\rangle \right]  \notag \\
&&\otimes |0\rangle _{L}|0\rangle _{R}.
\end{eqnarray}%
For $\frac{\Omega _{1}}{2}\tau _{2}=\pi /2$ and $\phi _{1}=3\pi /2$, the
state $-i|f\rangle |0\rangle _{L}|0\rangle _{R}$\ becomes $i|r\rangle
|0\rangle _{L}|0\rangle _{R}$.

\textit{Step~(iii)}: The system evolves for $\tau _{3}$ under $%
H_{3}=h_{L}\left( a_{L}|r\rangle \langle g|+\text{H.c.}\right) $. Then $%
U_{3}=\exp \left( -iH_{3}\tau _{3}\right) $ transforms
\begin{eqnarray}
|r\rangle |0\rangle _{L}|0\rangle _{R} &\rightarrow &\cos (h_{L}\tau
_{3})|r\rangle |0\rangle _{L}|0\rangle _{R}  \notag \\
&&-i\sin (h_{L}\tau _{3})|g\rangle |1\rangle _{L}|0\rangle _{R}.
\end{eqnarray}%
With $h_{L}\tau _{3}=\pi /2$, the state $i|r\rangle |0\rangle _{L}|0\rangle
_{R}$ becomes $|g\rangle |1\rangle _{L}|0\rangle _{R}$. Note that the state $%
|g\rangle |0\rangle _{L}|1\rangle _{R}$\ remains unchanged during the entire
operation. Hence, the three-step operations above complete the swap
operation in Eq.~($17$).

Compared to the use of five-level atoms, employing four-level atoms can
reduce the use of the pulses. However, one would have to prepare the initial
state $|0\rangle _{p}$~(i.e., $|0\rangle _{L}|1\rangle _{R}$) for each
cavity using auxiliary atoms, and the atoms would have to be moved out of
the cavities immediately after the swap operation, since the prepared photon
cluster states would otherwise change.

The swap operation in Eq.~($17$) can also be implemented using three-level
atoms~(the smallest number of levels needed for this approach). The
three-level case would have the disadvantages above for using four-level
atoms, and also the extra problem that adjusting the frequencies of the
cavity modes would be required. Thus, five-level atoms would be the optimal
choice based on the encoding of photonic qubits above. We note that there
exist proposals~(e.g.,~\cite{lim}) in which logical states of photonic
qubits are encoded as $|0\rangle $~(no photons) and $|1\rangle $~(one
photons). Compared with this type of encoding, encoding photon qubits with
polarization mode states~(in our work) can be used to perform single-qubit
rotations by passive linear optical elements~\cite{klm}, and thus it could
be used for optical QC.
\begin{figure}[tbp]
\includegraphics[width=7.0cm, height=3.8cm]{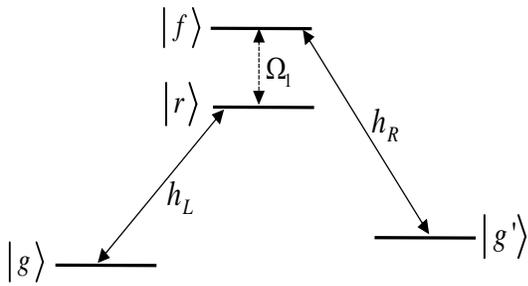}
\caption{Energy diagram of an atom with four levels. The transitions $%
|g\rangle \leftrightarrow |r\rangle $ and $|g^{\prime }\rangle
\leftrightarrow |f\rangle $ are resonantly coupled to the left~($\protect%
\sigma _{L}^{-}$) and the right~($\protect\sigma _{R}^{+}$) circularly
polarized mode of the cavity with coupling strengths $h_{L}$ and $h_{R}$,
respectively; while the transition $|r\rangle \leftrightarrow |f\rangle $ is
driven by a classical laser field with Rabi frequency $\Omega _{1}$.}
\end{figure}

\section{Concluding remarks}

\label{con}

In this work, we proposed an efficient and deterministic approach for
generating scalable cluster states for optical one-way QC in a photonic
system, by transferring the cluster state originally prepared in an atomic
system through unitary transformations. In this proposal, we can also
transfer part of the atomic cluster state to photons, e.g., transfer an
atomic single-qubit state of the atomic cluster state to a photonic qubit at
a time, for performing a single-qubit unitary operation and then
measurement~(readout).

The advantage of this approach is that the remaining part of the atomic
cluster state~(not transferred) is always stored in the atoms. This is a
hybrid way for robust one-way QC exploiting the advantages of both atomic
and photonic qubits: using atomic qubits for creating and storing a large
scale cluster state and also photonic qubits for performing single-qubit
rotations and measurements.

The first step is to make a large scale cluster state on atoms~($n$ qubits),
which can be easily generated~\cite{man} with atoms, and stored in atoms for
a long time. To do one-way optical QC, one would need to perform
single-qubit rotations and measurements on, say, the $i$th photonic qubit.
This would require swapping the $i$th atomic single-qubit state into a
photonic qubit, and measure the photonic qubit after rotating its state~(see
Fig.~$3(b)$). This swapping can be done deterministically using the approach
described above.

Finally, the state-transfer procedures introduced above can also be applied
in quantum communications~(e.g., transferring information between different
physical systems, trying to use advantages of both), storage of quantum
information, and quantum error correction.

\begin{acknowledgements}
We thank X.B.~Wang, L.-A.~Wu and S.B.~Zheng for helpful comments.
We acknowledge partial support from the National Security
Agency~(NSA), Laboratory of Physical Sciences~(LPS), Army Research
Office~(ARO), National Science Foundation~(NSF) Grant No.~0726909,
JSPS-RFBR Contract No.~09-02-92114, MEXT Kakenhi on Quantum
Cybernetics, and FIRST~(Funding Program for Innovative R\&D on
S\&T).
\end{acknowledgements}

\end{document}